\documentclass[prb,preprint,amsmath,amssymb,superscriptaddress,notitlepage]{revtex4-2}
\usepackage{epsfig}
\usepackage{graphicx}
\usepackage{color}
\usepackage{soul}
\usepackage{bm}
\usepackage{fixme}
\usepackage[colorinlistoftodos,textwidth=2.25cm]{todonotes} 
\usepackage[normalem]{ulem}
\usepackage[utf8]{inputenc}

\usepackage{titlesec}

\titleformat{\section}
{\normalfont\normalsize\bfseries} 
{\thesection}{1em}{} 
\renewcommand{\thesection}{\arabic{section}.}

\usepackage[left]{lineno}
\usepackage[normalem]{ulem}
\usepackage{soul}       
\usepackage[normalem]{ulem}
\usepackage{xcolor}     
\definecolor{lightyellow}{RGB}{255,255,150}
\definecolor{lightcyan}{RGB}{180,230,255}
\definecolor{lightgreen}{RGB}{200,255,200}
\definecolor{myblue}{RGB}{0,140,200}
\usepackage[
breaklinks=true,
pdfusetitle,
colorlinks=true,    
linkcolor=blue,
urlcolor=blue,
citecolor=blue,
filecolor=blue
]{hyperref}

\newcommand{\figref}[2]{\hyperref[#1]{Fig.~\ref*{#1}#2}}

\begin{document}

\title{Spin-orbit torque-driven synthetic antiferromagnetic oscillator}
\author{P.~K.~Rout}
\email{Pradeep-Kumar.Rout@physik.uni-regensburg.de}
\affiliation{Institute of Experimental and Applied Physics, University of  Regensburg, Universit\"asstrasse 31, 93051 Regensburg, Germany}
\author{J.~Godinho}
\affiliation{Institute of Experimental and Applied Physics, University of  Regensburg, Universit\"asstrasse 31, 93051 Regensburg, Germany}
\author{F.~Vilsmeier}
\affiliation{Fakult{\"a}t fur Physik, Technische Universit{\"a}t M\"unchen, 85748 Garching, Germany}
\author{R.~Salikhov}
\affiliation{Institute of Ion Beam Physics and Materials Research, Helmholtz-Zentrum Dresden-Rossendorf, Bautzner Landstrasse 400, 01328 Dresden, Germany}
\author{J. A. V\'elez}
\affiliation{Donostia International Physics Center, 20018 San Sebasti\'an, Spain}
\affiliation{Polymers and Advanced Materials Department: Physics, Chemistry, and Technology, University of the Basque Country, UPV/EHU, 20018 San Sebasti\'an, Spain}
\author{Z.~\v{S}ob\'a\v{n}}
\affiliation{Institute of Physics, Czech Academy of Sciences, Cukrovarnick\'a 10, 162 00, Praha 6, Czech Republic}
\author{D. Laroze}
\affiliation{ Instituto de Alta Investigaci\'{o}n, Universidad de Tarapac\'{a} , Casilla 7D, 1000000, Arica, Chile.}
\author{O. Gomonay}
\affiliation{Institute of Physics, Johannes Gutenberg-University Mainz, 55128 Mainz, Germany.}
\author{R.~M.~Otxoa}
\affiliation{Hitachi Cambridge Laboratory, Cambridge CB3 0HE, United Kingdom}
\affiliation{Donostia International Physics Center, 20018 San Sebasti\'an, Spain}
\author{C.~H.~Back}
\affiliation{Fakult{\"a}t fur Physik, Technische Universit{\"a}t M\"unchen, 85748 Garching, Germany}
\author{O.~Hellwig}
\affiliation{Institute of Ion Beam Physics and Materials Research, Helmholtz-Zentrum Dresden-Rossendorf, Bautzner Landstrasse 400, 01328 Dresden, Germany}
\affiliation{Institute of Physics, Chemnitz University of Technology, Reichenhainer Strasse 70, 09107 Chemnitz, Germany}
\author{J.~Wunderlich}
\email{Joerg.Wunderlich@physik.uni-regensburg.de}
\affiliation{Institute of Experimental and Applied Physics, University of  Regensburg, Universit\"asstrasse 31, 93051 Regensburg, Germany}
\affiliation{Institute of Physics, Czech Academy of Sciences, Cukrovarnick\'a 10, 162 00, Praha 6, Czech Republic}

\maketitle

{\bf

\noindent{Antiferromagnets offer a promising route toward robust spintronic devices because of their compensated magnetic order and exchange-enhanced spin dynamics. Here, we demonstrate a spin-orbit torque (SOT)-driven antiferromagnetic oscillator based on a nanoconstriction patterned from a synthetic antiferromagnet (SAF). Spin-rectification spectroscopy reveals electrical excitation of both acoustic and optical SAF eigenmodes, whose field and frequency dependences are quantitatively described by an antiferromagnetic resonance model. In addition to these linear eigenmodes, we observe low-field spin-rectification peaks that emerge only above a threshold DC current near the spin-flop transition. Their current-polarity-dependent sign and locking to an injected RF frequency provide electrical spin-rectification signatures consistent with current-selected chiral self-oscillatory dynamics. Micromagnetic simulations reproduce the threshold excitation of SOT-driven self-oscillations and injection locking, while macrospin simulations predict stable and chaotic nonlinear dynamics within the same spin-flop region. We interpret the multi-peak, weakly RF-frequency-dependent responses as a qualitative signature of complex nonlinear dynamics. These results establish SAF nanoconstrictions as an experimentally accessible platform for studying current-driven antiferromagnetic-like oscillator dynamics and motivate future work on nonlinear spintronic devices for signal processing and reservoir-computing concepts.}
}

\bigskip
\clearpage


\noindent{Spin-transfer torque (STT) and spin-orbit torque (SOT) nano-oscillators are at the forefront of spintronics research due to their potential applications in neuromorphic computing, microwave-assisted magnetic recording, magnonics, and wireless communications \cite{Torrejon2017, Grollier2020, Locatelli2014, Romera2018, Zahedinejad2020, Chen2016}. They have attracted considerable attention for their flexible device geometry, CMOS compatibility and robust mutual synchronization \cite{Villard2010, Chen2015, Romera2018, Zahedinejad2020}. STT nano-oscillators typically employ nanoscale spin valves or magnetic tunnel junctions, in which spin-polarized currents compensate magnetic damping to sustain microwave oscillations. Their frequencies are primarily determined by the applied magnetic field magnitude, and their inherently right-handed chirality is locked to the magnetic field orientation \cite{Slonczewski1996, Berger1996, Tsoi1998, Klselev2003, Demidov2007, Deac2008, Slavin2009}. Similarly, SOT nano-oscillators drive magnetization precession through charge-to-spin current conversion \cite{Demidov2012, Liu2012, Liu2013, Demidov2014, Duan2014, Haidar2019}. When based on ferromagnets, both STT and SOT oscillators exhibit limited controllability: at a fixed applied magnetic field, their right-handed chirality remains fixed, and their frequency can only be tuned slightly e.g., via voltage-controlled magnetic anisotropy \cite{Fulara2020, Choi2022}.}

Very promising alternative concepts for spintronic oscillators based on antiferromagnets have been theoretically proposed, combining ultrafast, exchange-enhanced spin dynamics that push oscillator frequencies into the THz regime with bias-current control over both frequency and chirality \cite{Cheng2016, Troncoso2019, Wolba2021, Parthasarathy2021, Gomonay2018, Han2023, Jungwirth2016}. This includes chiral self-oscillations as well as resonantly driven left- and right-handed circularly polarized modes in easy-axis systems, and linearly polarized resonant acoustic and optical modes in easy-plane antiferromagnets \cite{Gomonay2018, Han2023, Williamson1964, Keffer1952}.
Moreover, the dynamical range for sustaining self-oscillations is predicted to be significantly broader than in ferromagnetic oscillators \cite{Khymyn2018}. For example, chiral self-sustained N\'eel-order oscillations spanning the entire frequency range between the acoustic and optical eigenmodes have been predicted when an insulating antiferromagnet is excited by spin currents generated in an adjacent nonmagnetic metal layer \cite{Cheng2016}. One particularly compelling concept envisions SOT-driven self-oscillations directly in a metallic parity-time reversal ($\mathcal{PT}$)-symmetric antiferromagnet \cite{Troncoso2019}. In these conductive systems, self-oscillations and their chirality can be detected via phase-sensitive nonlinear rectification, arising from time-dependent magnetotransport phenomena such as anisotropic magnetoresistance (AMR) or the anomalous Hall effect (AHE) as found in noncollinear antiferromagnets \cite{HARDER2016, Sakamoto2024}.

In addition to periodic excitations, theoretical studies predict that spin torque-driven antiferromagnetic oscillators can exhibit complex nonlinear and chaotic regimes near the spin-flop transition \cite{Wolba2021, Parthasarathy2021}. Such regimes are of interest for nonlinear signal processing and reservoir-computing concepts \cite{Choi2019, Markovic2019}, although their experimental identification requires care because multi-peak or broadband spectra can also arise from nonlinear mode coupling or device-specific inhomogeneities. To our knowledge, experimental signatures consistent with current-driven self-oscillatory dynamics in antiferromagnetic systems have not been reported previously. In the present work, we therefore distinguish between the directly measured spin-rectification signatures and the simulation-supported interpretation of possible chaotic dynamics.

In this work, we investigate both resonant eigenmodes and self-sustained oscillations in a synthetic antiferromagnet (SAF) multilayer, which serves as a highly simplified yet versatile antiferromagnetic system. Despite its structural simplicity, the SAF retains all the essential ingredients required to support complex current-driven antiferromagnetic dynamics. Our SAF consists of two ferromagnetic layers separated by a nonmagnetic spacer that mediates a Ruderman-Kittel-Kasuya-Yosida (RKKY) interaction, resulting in antiferromagnetic coupling between the layers \cite{Duine2018}. Furthermore, the SAF is sandwiched between a tantalum (Ta) seed and platinum (Pt) cap layer, which possess opposite spin Hall angles. When a charge current flows through the multilayer stack, the heavy-metal interfaces generate spin currents that are injected into the SAF from both sides with identical spin polarization (\figref{fig_1}{a}) \cite{DYAKONOV1971, Hirsch1999, Wunderlich2005}. The resulting SOT acts on the antiferromagnetic order of the SAF in a manner analogous to the theoretically studied STT in collinear antiferromagnets, where chaotic self-oscillatory dynamics are predicted near the spin-flop transition \cite{Wolba2021}.

By driving the SAF with current-induced SOT, we demonstrate electrical excitation and detection of antiferromagnetic eigenmodes and investigate additional DC-current-induced nonlinear dynamics near the spin-flop transition. Using a combination of radio-frequency (RF) current excitation and an in-plane magnetic field with tunable magnitude and orientation, we detect linearly polarized acoustic and optical resonance modes through the spin rectification effect (SRE) \cite{Fang2011, Liu2011, HARDER2016}. Additional rectification peaks appear in the magnetic field dependence of the SRE signal when the applied DC current exceeds a critical threshold. These additional low-field SRE peaks appear in a regime where no zero-bias eigenmode resonance is detected at the probing frequency. We interpret these peaks as SRE signatures consistent with injection-locked, current-driven chiral self-oscillatory dynamics of the SAF order. Their amplitude increases monotonically with DC-current above the threshold, and their sign is governed by the DC-current polarity rather than by the magnetic-field polarity, consistent with current-selected chirality.

This interpretation is supported by micromagnetic simulations of threshold excitation of self-oscillations and injection locking. Complementary macrospin simulations predict stable and chaotic nonlinear regimes near the spin-flop transition through Lyapunov-exponent analysis and bifurcation diagrams. We note that direct detection of emission originated from self-oscillation, for example by Brillouin light scattering spectroscopy or spectrum-analyzer measurements, is not part of the present work; the DC-induced peaks are therefore discussed as SRE-based electrical signatures consistent with self-oscillatory dynamics and with complex nonlinear behavior.

Within this framework, injection locking provides a homodyne mechanism for electrically detecting DC-driven self-oscillatory dynamics: when the current-driven dynamics acquire spectral weight near the RF probe frequency and become phase-locked to the RF current, the time-dependent AMR/GMR mixes with the RF current to generate a rectified voltage. We therefore use this injection-locked SRE response as an electrical signature of DC-current-driven oscillatory dynamics.

\section{\bf Spin-orbit torque-driven excitation of SAF eigenmodes}

\noindent{To probe the spin dynamics of the SAF nanoconstriction, we applied a RF current and measured the resulting spin rectification voltage $V_{SRE}$ while sweeping an in-plane magnetic field. This voltage originates from nonlinear mixing of the time-dependent resistance, modulated by anisotropic magnetoresistance (AMR) and giant magnetoresistance (GMR), and the RF current. The rectification generates a second-order DC signal $V_{SRE}$, along with a high-frequency component at $2\omega_{\mathrm{RF}}$ \cite{Fang2011, Liu2011, HARDER2016}.}
	
The SAF multilayer stack (\figref{fig_1}{a}) comprises two 2-nm-thick permalloy ($\mathrm{Py: Ni_{0.81}Fe_{0.19}}$) layers, antiferromagnetically coupled through a 0.34-nm Ir spacer via the RKKY interaction. To enhance current density and spatially confine the magnetic excitations, the stack was patterned into a bar featuring a single notch-shaped nanoconstriction (\figref{fig_1}{b}). Vibrating sample magnetometry (VSM) reveals a predominant easy-plane magnetic anisotropy (Supplementary Section 1). An additional weak in-plane biaxial anisotropy is induced by the nanofabrication, which triggers a low-field spin-flop transition and is crucial for the emergence of DC-current-driven self-oscillations (Supplementary Section 2).

Applying an in-plane magnetic field induces a net magnetization ($\vec{M_{1}} + \vec{M_{2}}$) aligned with the field, and a N\'eel vector ($\vec{M_{1}} - \vec{M_{2}}$) oriented perpendicular to it. In the acoustic mode, the field-induced net magnetization precesses with constant magnitude around the direction of the applied field, whereas in the optical mode, the induced magnetization periodically varies its magnitude while maintaining its orientation along the field. As illustrated in \figref{fig_1}{c}, the acoustic and optical modes can be selectively or jointly excited depending on $\Phi$, consistent with the symmetry of the damping-like SOT excitation. \figref{fig_1}{d} shows distinct resonance peaks corresponding to the optical and acoustic eigenmodes of the SAF at 8 GHz, which arise from in-plane and out-of-plane oscillation of the net-magnetization, respectively. These excitations are driven by SOT generated at the Pt (top) and Ta (bottom) interfaces via the spin Hall effect. Due to their opposite spin Hall angles, both heavy-metal layers inject spin currents of identical polarization into the SAF from opposite sides. When the field is applied parallel to the current ($\Phi = 0^{\circ}$), only the acoustic mode is excited, as seen by the resonance peak at positive field polarity. The absence of a corresponding negative-field acoustic resonance is attributed to a pinned $180^{\circ}$ domain wall at the nanoconstriction, which can lead to cancellation of SRE contributions from oppositely oriented SAF domains (Supplementary Section 3). A slight misalignment (e.g., $\Phi = 5^{\circ}$) restores acoustic resonances at both field polarities, indicating a uniform magnetic state. At $\Phi = 45^{\circ}$, both acoustic and optical modes are observed. For $\Phi = 90^{\circ}$, only the optical mode is excited. The field-polarity-dependent sign reversal of both eigenmode SRE peaks is consistent with damping-like SOT as the dominant excitation mechanism \cite{Liu2011}, although field reversal alone does not provide a unique fingerprint of damping-like torque. Instead, the assignment follows from the combined mode selectivity, SAF symmetry, local domain configuration, and AMR/GMR rectification symmetry. For the optical mode, the antiphase motion of the two Py layers produces a dominant GMR-type resistance modulation, while the device symmetry strongly favors excitation by damping-like SOT; field-like and Oersted-field torques are expected to couple only weakly to this mode. For the acoustic mode, mainly detected through AMR-type rectification, field-like or Oersted excitation would produce a different field-polarity dependence from that observed upon reversal of the local SAF domain configuration. The observed behavior is therefore consistent with damping-like SOT as the dominant drive, although small field-like or Oersted contributions cannot be excluded solely from field reversal. 

A compact description of the rectification channels is given by $\Delta R(t) = \Delta R_{\mathrm{AMR}}(t) + \Delta R_{\mathrm{GMR}}(t)$, with $\Delta R_{\mathrm{AMR}} \propto \Delta[(\vec{M}_i \cdot \hat{J})^2]$ ($i = 1,2$ corresponding to the two FM layers of the SAF) and $\Delta R_{\mathrm{GMR}} \propto \Delta(\vec{M}_1 \cdot \vec{M}_2)$. For the acoustic mode, the two-layer magnetizations precess approximately in phase, so $\Delta(\vec{M}_1 \cdot \vec{M}_2)$ is very small and the GMR contribution is suppressed, while the AMR contribution remains finite. For the optical mode, the antiphase dynamics strongly modulate $\vec{M}_1 \cdot \vec{M}_2$, leading to a dominant GMR-type rectification signal. This contrast accounts for the opposite peak polarities of the acoustic and optical modes in \figref{fig_1}{d}. A full derivation of the AMR- and GMR-dominated spin-rectification responses for the present SAF geometry is provided in Supplementary Section~10.

At higher field magnitudes, Joule heating can introduce additional components to the detected signal through magneto-thermoelectric effects. In particular, a background voltage offset arises from the anomalous Nernst effect (ANE) or longitudinal spin Seebeck effect (SSE), which is driven by a vertical temperature gradient combined with a net magnetization component perpendicular to the bar. This net magnetization appears when the applied magnetic field tilts the two antiferromagnetically coupled layers such that a perpendicular component develops (see \figref{fig_1}{d} for $\Phi = 90^{\circ}$ and $\Phi = 45^{\circ}$, and Supplementary Section 4). The low-field DC-induced peaks discussed below (cf. Section 3) are measured in the $\vec{H} \parallel \vec{J}$ geometry, where the ANE contribution is symmetry-suppressed at the operating point. In addition, purely Joule-heating-driven rectification would produce signals that are even in DC current polarity. These symmetry considerations are relevant for excluding thermal contributions in the analysis that follows.

\figref{fig_1}{e} shows the frequency dependence of the acoustic mode: the resonance shifts to higher magnetic fields as the RF frequency increases, in agreement with the theoretical expectation. In \figref{fig_1}{f}, we compare the experimental resonance fields with the analytical expressions for easy-plane antiferromagnets. The acoustic mode follows ${f_{acoustic}} = {\gamma}H(1+|H_{an}|/{2H_{ex}})^{1/2}$, and the optical mode ${f_{optical}} = {\gamma}\{2H_{ex}|H_{an}|$ - $(|H_{an}|/2H_{ex}) H^{2}\}^{1/2}$ where $H_{ex}$ and $H_{an}$ corresponds to the antiferromagnetic RKKY exchange field and anisotropy field respectively. Using independently determined values $\mu_{0}H_{ex}$ = 0.25 T and $\mu_{0}H_{an}$ = 0.98 T from VSM measurements (Supplementary Section 1), we find excellent agreement between experiment and theory.

\section{\bf Micromagnetic simulation of chiral self-oscillations}

\noindent
To gain further insight into the current-driven spin dynamics observed in the SAF nanoconstriction, we performed micromagnetic simulations using the GPU-accelerated open-source solver MuMax3 (see Methods) \cite{Vansteenkiste2014}. While micromagnetic simulations neglect exchange damping, an effect that becomes relevant in true antiferromagnets with strong atomic-scale exchange interactions, they remain well-suited for studying SAF systems composed of ferromagnetic layers. In such multilayers, the local magnetizations are antiferromagnetically coupled only through the comparatively weak RKKY interaction across a non-magnetic spacer, which justifies treating their dynamics within a conventional micromagnetic framework \cite{Dai2021, Wang2024, YangLiu2024}.

The simulations incorporate the experimentally determined material parameters and geometry of the SAF nanoconstriction shown in \figref{fig_1}{}. They reveal the emergence of self-oscillatory dynamics when the applied DC current exceeds a critical threshold. To characterize the antiferromagnetic dynamics, we analyze the time evolution of the net magnetization ($\vec{M} = \vec{M_{1}} + \vec{M_{2}}$), which becomes non-zero only when the magnetizations of the two layers are dynamically canted. Such a canting arises during antiferromagnetic excitations and during the propagation of antiferromagnetic magnons.

\figref{fig_2}{} presents representative simulation results illustrating the behavior of the SAF oscillator in different driving regimes at an applied magnetic field of 45 mT along the bar direction. When a subthreshold DC current corresponding to an average current density of $J_{dc}^{below}$ = $1.7 \times 10^{8}~A/cm^{2}$ at the nanoconstriction, is applied, the damping-like SOT slightly tilts the layer magnetizations along the out-of-plane direction, resulting in a weak and static net magnetization in the constriction (\figref{fig_2}{a(i)}). The corresponding power spectral density (PSD), computed from the magnetization dynamics in the central nanoconstriction, exhibits no distinct spectral features (\figref{fig_2}{b(i)}), indicating the absence of self-oscillations.

In contrast, applying a superthreshold DC current density of $J_{dc}^{above}$ = $2.2 \times 10^{8}~A/cm^{2}$ induces persistent excitations localized in the nanoconstriction region (\figref{fig_2}{a(ii)}). The associated PSD (Fig.~2b(ii)) shows a broad frequency distribution with an amplitude enhancement near the optical and acoustic eigenmode frequencies of the SAF. Such enhancement is expected, as self-sustained oscillations typically manifest through the excitation of the system's intrinsic eigenmodes. Importantly, unlike the linear optical mode where antiphase precession cancels the net magnetization, the nonlinear self-oscillatory regime produces a finite, oscillatory net $m_z$ component visible in the simulated PSD. Correspondingly, these excitations emit propagating spin waves with short wavelengths ($\leq$ 50 nm) radiating from the constriction into the surrounding SAF region. 

When an additional radio-frequency current density of $J_{RF}^{peak}$ = $0.5 \times 10^{8}~A/cm^{2}$ at 8 GHz is applied, a portion of the self-oscillatory motion becomes phase-locked to the additional monochromatic RF excitation, as shown in \figref{fig_2}{a(iii)}. This phenomenon, known as injection-locking, leads to a pronounced modification of the spin-wave pattern: longer-wavelength magnons, consistent with the locking frequency of the RF current, dominate the propagation away from the constriction. The corresponding PSD (\figref{fig_2}{b(iii)}) reveals a sharp peak at the locking frequency, confirming synchronization between the DC driven self-oscillation and the RF current. In the Supplementary Material, we provide two animations in which the antiferromagnetic dynamics and magnon generation due to self-oscillations can be traced without and with injection locking at 8 GHz locking frequency. While our main results focus on the experimentally accessible 8 GHz range, the Supplementary Section 5 demonstrates that injection locking is not limited to this frequency; we show consistent locking across a range of 5-8 GHz and at various magnetic fields (30-50 mT) near the spin-flop transition. Importantly, injection locking provides a mechanism for detecting DC-induced self-oscillations via spin rectification, as the synchronized precession component contributes a coherent oscillatory resistance at the RF frequency. This homodyne detection mechanism requires that the spectrum of the SOT-induced excitation contains the RF frequency $f_{1}$ and that the precession remains phase-locked to the RF current \cite{Sakamoto2024}.

\section{\bf Threshold current and SRE signatures of chiral self-oscillatory dynamics}

We now discuss SRE signatures of DC-current-driven non-linear dynamics near the spin-flop transition, i.e., at low applied magnetic fields ($|\mu_{0}H| < 50$ mT), where no resonant eigenmodes are observed in the absence of a DC current [Supplementary Section 2]. Although the external magnetic field is applied parallel to the charge-current direction in the measurements of \figref{fig_3}{}, the low-field operating point lies near the spin-flop transition of the patterned SAF, where the two sublayer magnetizations become canted away from perfect collinearity. In this regime, the damping-like torque $\vec{\tau}_{\mathrm{DL}} \propto \vec{M}_i \times (\vec{M}_i \times \vec{\sigma})$ is finite on each layer. Because the two layers are antiferromagnetically coupled and receive spin currents from opposite interfaces with the same spin polarization, the resulting torques act as an effective staggered antidamping drive on the SAF order. This provides a mechanism by which DC current can compensate damping and excite oscillatory dynamics even in the $\vec{H} \parallel \vec{J}$ geometry. \figref{fig_3}{a} shows the spin rectification voltage $V_{SRE}$ measured as a function of the applied in-plane magnetic field ($\Phi = 0^{\circ}$) for three cases: zero DC current ($I_{dc} = 0$ mA) and finite DC currents of $I_{dc} = \pm 0.45$ mA. The corresponding current density within the nanoconstriction reaches $2.9 \times 10^{7}~A/cm^{2}$. At zero DC current, no resonant features are observed in the SRE signal at the RF probing frequency of 8 GHz, consistent with the absence of both SOT-driven eigenmode excitation and self-oscillations. In contrast, for both positive and negative DC super threshold currents, sharp SRE peaks emerge at distinct magnetic field values, consistent with the onset of injection-locked DC-driven self-oscillatory dynamics. Strikingly, the sign of these DC current induced SRE peaks is governed solely by the polarity of the applied DC current and remains independent of the magnetic-field direction, a behavior fundamentally distinct from the conventional field-polarity-dependent SRE response of linear eigenmodes observed in the absence of DC bias (cf. Section~1).

We attribute these threshold dependent resonant peaks to self-sustained oscillations driven by the DC current-induced damping-like SOT, which become injection-locked to the RF frequency. The observed current-polarity-dependent SRE signal supports the interpretation of current-selected chirality. Unlike the linearly polarized eigenmodes, these DC-induced self-oscillations emerge near the spin-flop transition, a region where the SAF is highly susceptible to SOT-driven instabilities \cite{Wolba2021}. In the vicinity of the spin-flop transition, the system's nonlinear dynamical response is dictated predominantly by the current-induced SOT. In our SAF geometry, the damping-like SOT acts as a staggered effective field across the two antiferromagnetically coupled layers; this induces chiral self-oscillatory dynamics with a handedness determined by the current polarity. We define the handedness by the sense of rotation of the SAF order parameter projected onto the plane perpendicular to its equilibrium direction and viewed along the positive current direction. Reversing $I_{\mathrm{dc}}$ reverses the spin accumulation at the Pt and Ta interfaces and hence changes the sign of the damping-like SOT. In the simulations, this reverses the rotation sense of the DC-driven limit-cycle component and shifts the phase of the dynamic AMR/GMR response by approximately $\pi$, producing the observed reversal of $V_{\mathrm{SRE}}$. Because AMR and GMR are even under simultaneous reversal of both layer magnetizations, the sign of the low-field SRE peaks is insensitive to magnetic-field polarity but reverses with current polarity.

Our observations align with the theoretical predictions for in-plane antiferromagnetic spin Hall nano-oscillators, in which the chirality of the self-oscillations is determined by the direction of the applied current \cite{Cheng2016}. We note that SRE-based electrical detection has been demonstrated to be directly sensitive to magnon chirality in compensated ferri- and antiferromagnetic systems \cite{Wang2024_PRL, Shiota2024}, though those works address chiral linear eigenmodes rather than the DC-driven self-oscillatory states reported here, establishing that the sign of the rectified voltage reflects the handedness of the excited mode. A conceptually related phenomenon was recently reported in nearly isotropic ferromagnets \cite{Kurebayashi2026}, where SOT-driven dynamical instabilities stabilize the magnetization against the applied field, mimicking a spintronic Kapitza pendulum. In that system, the resulting nonlinear state is probed via SRE, which exhibits a torque-dependent sign reversal. This strengthens our observation of a current-polarity-governed SRE response and highlights the utility of spin-rectification spectroscopy in identifying handedness of the magnetization dynamics. Furthermore, using Fourier analysis of the time-dependent AMR and GMR responses in our micromagnetic simulations, we numerically confirm that the polarity of the SRE voltage arising from injection-locked self-oscillations is solely selected by the DC current polarity and independent of the applied magnetic field  (Supplementary Material section 6). We thus conclude that, in the low-field regime near the spin-flop transition, the chirality of the self-oscillations is consistent with current-polarity-selected handedness, an important distinction from ferromagnetic STT and SOT oscillators, where the chirality is always righthanded with respect to the applied magnetic field. We emphasize that current-polarity-dependent sign reversal of $V_\mathrm{SRE}$ can also occur in nonlinear rectification response of conventional ferromagnetic eigenmodes. The key distinction here is that the low-field peaks are absent at zero DC bias and emerge only above a threshold current within a field-frequency window where no conventional SAF eigenmode resonance is detected. We therefore do not interpret these peaks as a DC-bias modification of a pre-existing resonance, but rather as SRE signatures of a DC-induced dynamical state. The comparable amplitudes for opposite current polarities, together with the magnetic-field-polarity independence of the rectified-voltage sign, further support the interpretation in terms of current-selected chirality.

The threshold behavior of the self-oscillations is illustrated in \figref{fig_3}{b}. Focusing on a representative SRE peak within the narrow field range between –55 mT and –35 mT, we vary the DC current amplitude corresponding to current densities from 0 to $2.9 \times 10^{7}~A/cm^{2}$. The selected SRE peak appears only above a critical current density of $1.3 \times 10^{7}~A/cm^{2}$, and shifts slightly to lower field as the current is increased. The threshold current density in zero-temperature micromagnetic simulations is about one order of magnitude larger. We therefore do not regard the absolute threshold as quantitatively reproduced by the simulations. Instead, the simulations are used to identify the qualitative mechanism and symmetry of the instability. The difference may arise from thermal activation, Joule-heating-induced changes of material parameters, and uncertainties in the local current density/current partitioning within the multilayer. Moreover, the experimental current density quoted above refers to the estimated average current density in the nanoconstriction. It was calculated from the applied DC current and the effective conducting cross-section, taking into account the multilayer thickness and current partitioning obtained from COMSOL simulations. The uncertainty is dominated by the local current crowding at the notch, resistivity values of individual layers, and the exact constriction width. We therefore compare experiment and simulation primarily at the level of qualitative trends rather than absolute threshold values.

\section{\bf Current-driven dynamics close to the spin-flop transition with possible signatures of chaotic behavior}

\noindent{To investigate the complex dynamics predicted for SOT-driven antiferromagnetic oscillators near the spin-flop transition, we performed spin rectification measurements across a range of RF frequencies (5-8\,GHz) and compared the acoustic eigenmode of Section~1 with the DC-driven self-excitations described in Section~3. In the absence of a DC current, only the acoustic eigenmode is observed, with a resonance field that shifts monotonically to higher values as the RF frequency increases (\figref{fig_4}{a}), in agreement with the analytical model discussed in Section~1.}

When a DC current with a density exceeding the critical threshold ($J_{\mathrm{dc}}^{\mathrm{above}} = 2.5 \times 10^{7}\,\mathrm{A/cm}^{2}$) is applied, a series of additional sharp SRE peaks emerges near zero applied magnetic field, close to the spin-flop transition (\figref{fig_4}{d}). These peaks are absent under zero DC bias (\figref{fig_4}{c}) and are assigned to the injection-locked DC-driven dynamics discussed in Section~3. In contrast to the acoustic eigenmode (\figref{fig_4}{b}), whose resonance field shifts systematically with RF frequency, the dominant low-field peaks remain confined to a narrow field interval as the RF frequency is varied from 5 to 8~GHz. This frequency-independent field-degenerate behavior is consistent with injection locking to a broadband current-driven dynamical spectrum, as found in the micromagnetic simulation results of Section~2 (cf.,\ Fig.~2b (ii), (iii)). It is not, by itself, a definitive proof of chaos.

To further examine the nature of these excitations, we carried out macrospin simulations using experimentally determined parameters for the RKKY exchange coupling and magnetic anisotropy. While micromagnetic simulations resolve spatial features of the spin dynamics, macrospin simulations enable longer integration times and provide access to the global dynamical behavior over extended timescales and parameter space. The macrospin simulations confirm the emergence of complex and highly nonlinear magnetization dynamics near the spin-flop transition. In particular, Lyapunov exponent analysis and bifurcation diagrams (Supplementary Section~7) reveal both stable and chaotic oscillatory regimes, depending on the applied DC current and magnetic field. The experimental observation of multiple threshold-dependent SRE peaks with weak or no RF-frequency dependence is consistent with this predicted complex nonlinear regime. However, because the present measurements do not include real-time phase trajectories, fluctuation spectra, Brillouin light scattering, or direct microwave-emission measurements, we do not claim unambiguous experimental evidence of chaos. Instead, we regard the data as SRE-based qualitative evidence for complex nonlinear dynamics in the parameter region where the simulations predict a route to chaos.

Finally, we observe a subtle but systematic dependence of the resonance field of the 8\,GHz injection-locked mode on the magnitude of the DC current. As detailed in Supplementary Section~7, increasing the current results in a slight shift ($\approx 1$\,mT) of the resonance field, further highlighting the current-controlled nature of these excitations.

\section{\bf Conclusion}

\noindent{We have demonstrated electrical signatures of a SOT-driven synthetic antiferromagnetic oscillator based on a SAF nanoconstriction. Spin-rectification spectroscopy resolves the acoustic and optical antiferromagnetic-like eigenmodes and shows that their field and frequency dependence are quantitatively captured by an analytical antiferromagnetic resonance model. Beyond these linear modes, we observe additional low-field SRE peaks that emerge only above a threshold DC current near the spin-flop transition. Their current-polarity-dependent sign and RF injection-locking behavior provide electrical signatures consistent with current-driven chiral self-oscillatory dynamics of the SAF order.}

Micromagnetic simulations support this interpretation by reproducing threshold excitation, broadband spectral content, and injection locking to an RF probe current. Complementary macrospin simulations predict stable and chaotic nonlinear regimes within the same spin-flop region. In the absence of direct microwave-emission or Brillouin light-scattering measurements in the present work, the multi-peak, weakly RF-frequency-dependent response is interpreted as qualitative evidence for complex nonlinear dynamics rather than as definitive experimental evidence of chaos.

We note that the present electrical detection scheme, based on AMR and 
GMR signals of $\sim$0.1\%, inherently limits the absolute signal amplitude, a constraint shared by most metallic SOT nano-oscillators relying on AMR or spin Hall magnetoresistance (SMR) readout. Two complementary routes toward enhanced detection are identified for future work: three-terminal architectures incorporating a magnetic tunnel junction for TMR-based electrical readout, a strategy currently being explored for SAF-based systems \cite{Li2025}, and optical detection via Brillouin light scattering spectroscopy, which is not limited by magnetoresistive signal amplitudes and would provide direct spatial and spectral verification of the current-induced excitations within the nanoconstriction.

Together, these results establish SAF nanoconstrictions as a controllable platform for studying SOT-driven antiferromagnetic oscillator dynamics in the GHz regime, including current-selected chirality and nonlinear dynamics near the spin-flop transition, and motivate future studies of nonlinear spintronic signal-processing and reservoir-computing concepts.
\newpage
\noindent
{\bf Methods}

\noindent
{\bf Sample fabrication}

\noindent
For these studies, we prepared synthetic antiferromagnets (SAFs) consisting of two Py ($Ni_{81}Fe_{19}$) ferromagnetic layers exhibiting easy-plane anisotropy. The antiferromagnetic coupling between these layers was achieved via the Ruderman-Kittel-Kasuya-Yoshida (RKKY) interaction, through a 0.34 nm Ir spacer layer. A 5-nm Ta seedlayer was included underneath the stack to improve adhesion.  The whole Ta(5 nm)/Py(2 nm)/Ir(0.34 nm)/Py(2 nm)/Pt(5 nm) multilayer was deposited at room temperature by d.c. magnetron sputtering  in an ultrahigh-vacuum BESTEC system. Argon was supplied at a flow of 20 standard cubic centimeters per minute (sccm), resulting in an Ar atmosphere at 0.4 Pa (3 mTorr). The substrates were Si wafers with a 100-nm thermally oxidized SiO$_{2}$  layer. During deposition, the substrate was rotated at $\approx$ 60 rpm. The d.c. sputter power was set to 50 W for Py, Pt, and Ta, and 40 W for Ir corresponding to the growth rates of 0.255 nm/s for Py, 0.1 nm/s for Pt, 0.068 nm/s for Ta, and 0.056 nm/s for Ir. Layer thicknesses were controlled via deposition time, and sputter rates were calibrated in advance by X-ray reflectivity.

The constrictions of the devices were defined by electron-beam lithography (EBL) using a positive e-beam resist (AR-P 6200). The exposed regions were etched by argon ion milling in a reactive-ion etching-inductively coupled plasma (RIE-ICP) system. Electrical contacts were fabricated using a standard lift-off process following thermal evaporation of a Cr/Au bilayer (5\,nm/80\,nm).

\noindent
{\bf Spin rectification effect (SRE) measurements}

\noindent
All electrical SRE measurements were carried out at room temperature. For the SRE measurements, the chip containing the SAF nano-constriction device is mounted on a custom made RF-PCB with coplanar waveguide (CPW) for high frequency access. The SAF nano-constriction device was then bonded to the CPW. A radio frequency (RF) current ($I_{RF}$) and direct current ($I_{DC}$) were applied to the device via the high-frequency and low-frequency ports of a bias tee, respectively. For applying the RF current, we used a signal generator, and for the direct current, a precision source measure unit was used. The device, mounted on an RF PCB, was placed inside a room-temperature electromagnet, and the magnetic field was monitored with a Hall sensor and gaussmeter. For the spin rectification effect (SRE) measurements, we employed a field-sweep method, in which the magnetic field was swept at a fixed RF frequency to identify resonances. The RF current driven through the SAF nanoconstriction generates spin-orbit torque and Oersted-field torque, which can both in principle lead to excitation of the N\'eel order oscillating at the frequency of the RF current. However, in our system, current-induced Oersted-field torque contributions are expected to be small due to their calculated magnitude and symmetry. As can be seen from the Supplementary Fig. S12 the maximum $z$-component of the Oersted field at the notch is about $\approx$ 0.4 mT for $J_{dc}^{above}$ = $2.2 \times 10^{8}~A/cm^{2}$. The oscillating N\'eel order, in turn, produces a time-dependent change in the magnetoresistance (GMR and AMR) of the SAF, which mixes with the RF current to yield a rectified dc voltage $V_{SRE}$. Because  $V_{SRE}$  is very small, we used amplitude modulation of the RF current at 333 Hz and lock-in detection. The device with the RF PCB assembly was mounted on a non-magnetic piezo rotation stage to enable measurements at different in-plane angles.

\noindent
{\bf Micromagnetic simulations}

\noindent
Micromagnetic simulations were performed with MuMax3 \cite{Vansteenkiste2014} on reduced device dimensions of 1 $\mu$m $\times$ 0.5 $\mu$m with a 70 nm constriction width, scaled from the experimental 15 $\mu$m $\times$ 5 $\mu$m geometry, preserving the aspect ratios and therefore the asymmetric constriction with $90^{o}$ opening angle. Only the magnetization distribution of the two Py layers of the SAF were modeled using MuMax3. Current and Oersted-field distributions were obtained from COMSOL simulations of the same device geometry and used as input for the MuMax3 simulation (Supplementary Note 8). The SAF stack was discretized on a 512 $\times$ 256 $\times$ 3 mesh, corresponding to a cell size of about 2 $\times$ 2 $\times$ 1 $\mathrm{nm^{3}}$. To suppress artefacts from magnons reflected at the wide leads far from the constriction, Gilbert damping was gradually increased at the outer device regions along with the periodic boundary conditions to mimic the impact of the larger dimensions of the real device. Material parameters were: saturation magnetization $M_{s} = 7.9 \times 10^{5}$ A/m, intra-layer exchange stiffness $A = 1.1\times10^{-11}$ J/m and a Gilbert damping  $\alpha = 0.01$. The interlayer AF coupling was tuned to reproduce experimental spin-flip fields in both in-plane and out-of-plane geometries, giving 0.25 T for the critical field. A spin Hall angle of 0.15 was used to model the SOT. The system was first relaxed to the ground state, then evolved under current and/or field excitation until the steady-state regime was reached.

\noindent
{\bf Macrospin simulations}

\noindent
Macrospin simulations were performed by numerically solving coupled Landau--Lifshitz--Gilbert equations for the two ferromagnetic layers of the SAF, including interlayer antiferromagnetic exchange coupling as well as current induced SOTs. The effective field contained external, exchange, and anisotropy contributions. Material parameters were chosen to match the experimentally relevant regime and were consistent with the micromagnetic simulations, including a saturation magnetization $M_{s} = 7.9 \times 10^{5}$~A/m, Gilbert damping $\alpha = 0.01$, antiferromagnetic interlayer exchange field $\mu_{0}H_{\mathrm{exc}} = 250$~mT, and spin Hall angle $\theta_{\mathrm{SH}} = 0.15$. An effective out-of-plane hard-axis anisotropy of $\mu_{0}H_{u,z} \approx -1$~T was included to reproduce the easy-plane character originating from demagnetizing fields that are not explicitly treated within the macrospin approximation. Further, the simulations included an effective in-plane easy-axis anisotropy of $\mu_{0}H_{u,x} \approx 5$~mT to mimic the shape-induced anisotropy associated with the patterned device geometry, which arises naturally in the micromagnetic simulations.  The coupled equations were solved using a fourth-order Runge--Kutta method with fixed time step $\Delta \tau = 5 \times 10^{-3}$. Dynamical phase diagrams were constructed as functions of DC current density and external magnetic field. Fourier spectra, bifurcation diagrams, and Lyapunov-exponent analyses were used to identify stable, periodic, and chaotic nonlinear regimes. Additional details of the model, numerical implementation, and nonlinear-dynamics analysis are provided in Supplementary Section~7.

\noindent
{\bf Data availability}

\noindent
The data of this work are included in the article and it's Supplementary Information. Additional raw data are available from the corresponding authors upon request.

\noindent
{\bf Code availability}

\noindent
The micromagnetic and macrospin simulation codes are available from the corresponding authors upon reasonable request.


\noindent
{\bf Acknowledgements}

\noindent
This work was supported in part by the Deutsche Forschungsgemeinschaft (DFG, German Research Foundation) within Project-ID 314695032 – SFB 1277 “Emergente relativistische Effekte in der Kondensierten Materie”, Project-ID 452301518 “Investigation of quench switching of antiferromagnets with high spatial and temporal resolution”, and by the European Union’s Horizon 2020 research and innovation program under the Marie Skłodowska-Curie Grant Agreement No. 861300 “Cold Opto-Magnetism for Random Access Devices”. The work had also the support from the Czech Science Foundation within the Project GACR 21‑28876J. This work was also supported by DFG through Project No. 514946929. DL acknowledges partial financial support from FONDECYT 1240985 and FONDECYT 1231020 as well as from Centers of Excellence with BASAL/ANID financing, CIA250002, CEDENNA.

\noindent
{\bf Author contributions}

\noindent
P.K.R. and J.W. conceived the project. P.K.R., J.G., and J.W. designed the AF-oscillator devices. R.S. and O.H. grew and characterized the SAF multilayers. Z.S. fabricated the nanoconstriction devices. P.K.R., J.G., F.V., C.H.B., and J.W. designed and performed the SRE experiments. P.K.R., F.V., and C.H.B. designed and performed the TR-MOKE experiments. P.K.R. and J.G. carried out the micromagnetic simulations. J.A.V., D.L., and R.M.O. carried out the macrospin simulations. P.K.R., J.G., R.S., R.M.O., O.H, and J.W. discussed the relationship between the simulation and experimental results. P.K.R., J.G., and J.W. wrote the manuscript with input from all authors.  


\bibliographystyle{naturemag}
\bibliography{Refs}


\newpage

\noindent {\bf Figure Captions}
\vspace*{1cm}

\noindent {\bf Fig.~\ref{fig_1}}
{\bf Spin-orbit torque-driven excitation of acoustic and optical eigenmodes.} 

\noindent{{\bf a)}, Schematic of the SAF multilayer comprising two permalloy ($\mathrm{Py: Ni_{0.81}Fe_{0.19}}$) layers antiferromagnetically coupled via a 0.34-nm Ir spacer, sandwiched between Pt and Ta layers. The green arrows indicate the spin polarization directions injected from the heavy-metal interfaces when a charge current flows through the multilayer stack. Red and blue arrows represent the magnetization vectors ($\vec{M_{1}}$) and ($\vec{M_{2}}$) of the upper and lower Py layers, respectively. The in-plane magnetic field ($\vec{H}$) and charge current ($\vec{J}$) are applied at an angle $\Phi$. {\bf b)}, Scanning electron micrograph of a fabricated nanoconstriction device with a constriction width (w) of approximately 170 nm. {\bf c)}, Illustration of the magnetization dynamics for the acoustic mode (top, excited at $\Phi = 0^{o}$) and optical mode (bottom, excited at $\Phi = 90^{o}$). The acoustic mode corresponds to out-of-plane precession of the net magnetization ($\vec{M} = \vec{M_{1}} + \vec{M_{2}}$), while the optical mode corresponds to in-plane precession. {\bf d)}, Spin rectification voltage $V_{SRE}$ measured as a function of magnetic field for $\Phi = 0^{o}$, $45^{o}$, and $90^{o}$ at 8 GHz and 15 dBm RF power. Depending on $\Phi$, the data show either selective or combined excitation of acoustic and optical modes. {\bf e)}, $V_{SRE}$ versus magnetic field for different RF frequencies from 2 GHz to 12 GHz, showing the frequency dependence of the acoustic mode. {\bf f)}, Extracted resonance frequencies of the acoustic (red symbols) and optical (blue symbols) modes versus magnetic field. Solid and dashed black lines follow the analytical mode frequencies for easy-plane antiferromagnets using the values $\mu_{0}H_{ex}$ = 0.25 T and $\mu_{0}H_{an}$ = 0.98 T, determined independently from VSM measurements. The star corresponds to the measurements of (d) for 8 GHz.
	
\vspace*{1cm}

\noindent {\bf Fig.~\ref{fig_2}}
{\bf Micromagnetic simulations of SOT driven self-oscillatory dynamics.} 

\noindent{{\bf a)}, Time-snapshots of normalized net-magnetization at {\bf(i)} $J_{dc}^{below}$, {\bf(ii)} $J_{dc}^{above}$, and {\bf(iii)} $J_{dc}^{above}$ + $J^{RF}$ . {\bf b)}, Power spectral densities (PSD) at the three different bias conditions. ($J_{dc}^{below}$ = $1.7 \times 10^{8}~A/cm^{2}$, $J_{dc}^{above}$ = $2.2 \times 10^{8}~A/cm^{2}$; $J_{RF} = A sin(2 \pi f t)$ with f= 8 GHz, A= $0.5 \times 10^{8}~A/cm^{2}$; $m_{z} = \hat{z}.(\vec{M_{1}} + \vec{M_{2}})/(|\vec{M_{1}}| + |\vec{M_{2}}|)$}. During the simulations, a small magnetic field of 45 mT is applied along the bar direction.

\vspace*{1cm}

\noindent {\bf Fig.~\ref{fig_3}}
{\bf SRE signatures consistent with DC-current-induced self-oscillatory dynamics in SAF.} 

\noindent{{\bf a)}, Electrical measurements of the SRE signal for self-oscillations vs. applied magnetic field for three different dc current densities; $0$ (black), $-2.9 \times 10^{7} ~A/cm^{2}$ (blue) and $+2.9 \times 10^{7} ~A/cm^{2}$ (red) with RF current at 8 GHz and 15 dBm. The two chiralities of self-oscillation modes, left-handed mode (LHM) and right handed mode (RHM), are shown as schematic. LHM/RHM denote the handedness convention defined in the main text; the assignment is based on the current polarity and simulation phase of the dynamic AMR/GMR response. {\bf b)}, DC current density dependencies of a single self-oscillation mode from $0~A/cm^{2}$ to $2.9 \times 10^{7} ~A/cm^{2}$ showing the threshold behavior of self-oscillation. The magnetic field in this case is applied along the bar direction ($\Phi = 0^{o}$).}

\vspace*{1cm}

\noindent {\bf Fig.~\ref{fig_4}}
{\bf Frequency dependencies of the SRE signatures of DC-current-driven dynamics in a SAF.} 

\noindent{{\bf a)} and {\bf b)} show the acoustic modes for frequencies between 5 GHz to 8 GHz and with $J_{dc} = 0$ and $J_{dc} = +2.9 \times 10^{7} ~A/cm^{2}$ respectively obtained from the SRE measurements. The SRE measurements with the injection-locked self-oscillations for frequencies from 5 GHz to 8 GHz are shown in {\bf c)} and {\bf d)} for  $J_{dc} = 0$ and $J_{dc} = +2.9 \times 10^{7} ~A/cm^{2}$ respectively. In {\bf c)} and {\bf d)}, the measured SRE voltages are shifted along y-axis for different frequencies for better visualizations of the self-oscillation modes. The magnetic field is applied along the bar direction with a slight misalignment and the RF power for all the frequencies was kept fixed at 15 dBm.}

\vspace*{1cm}

\newpage
\begin{figure}[h!]
\hspace*{-0cm}
\includegraphics[scale = 0.95]{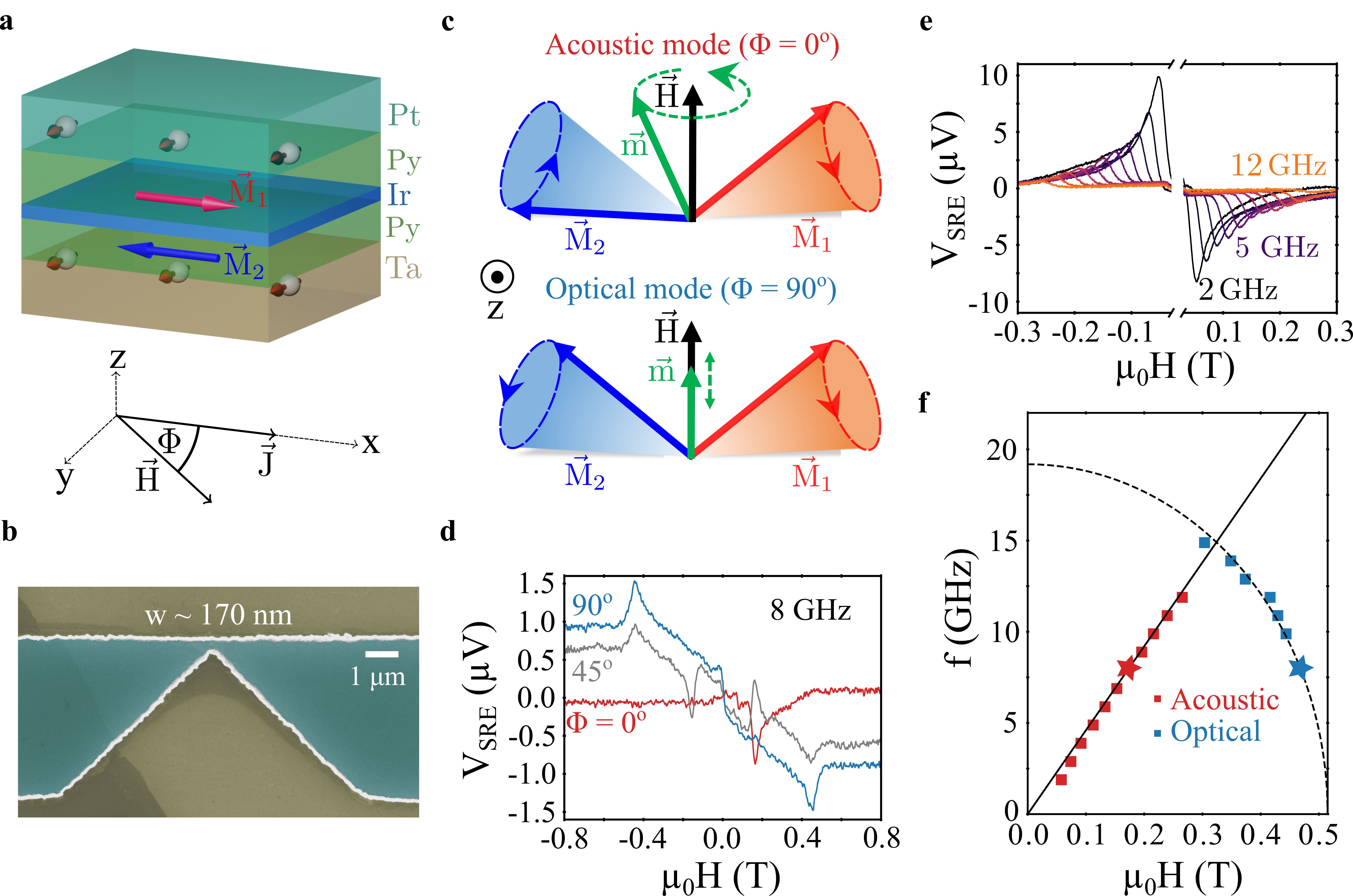}
\caption
{
}
\label{fig_1}
\end{figure}

\newpage
\begin{figure}[h!]
\hspace*{-0cm}
\includegraphics[scale = 0.95]{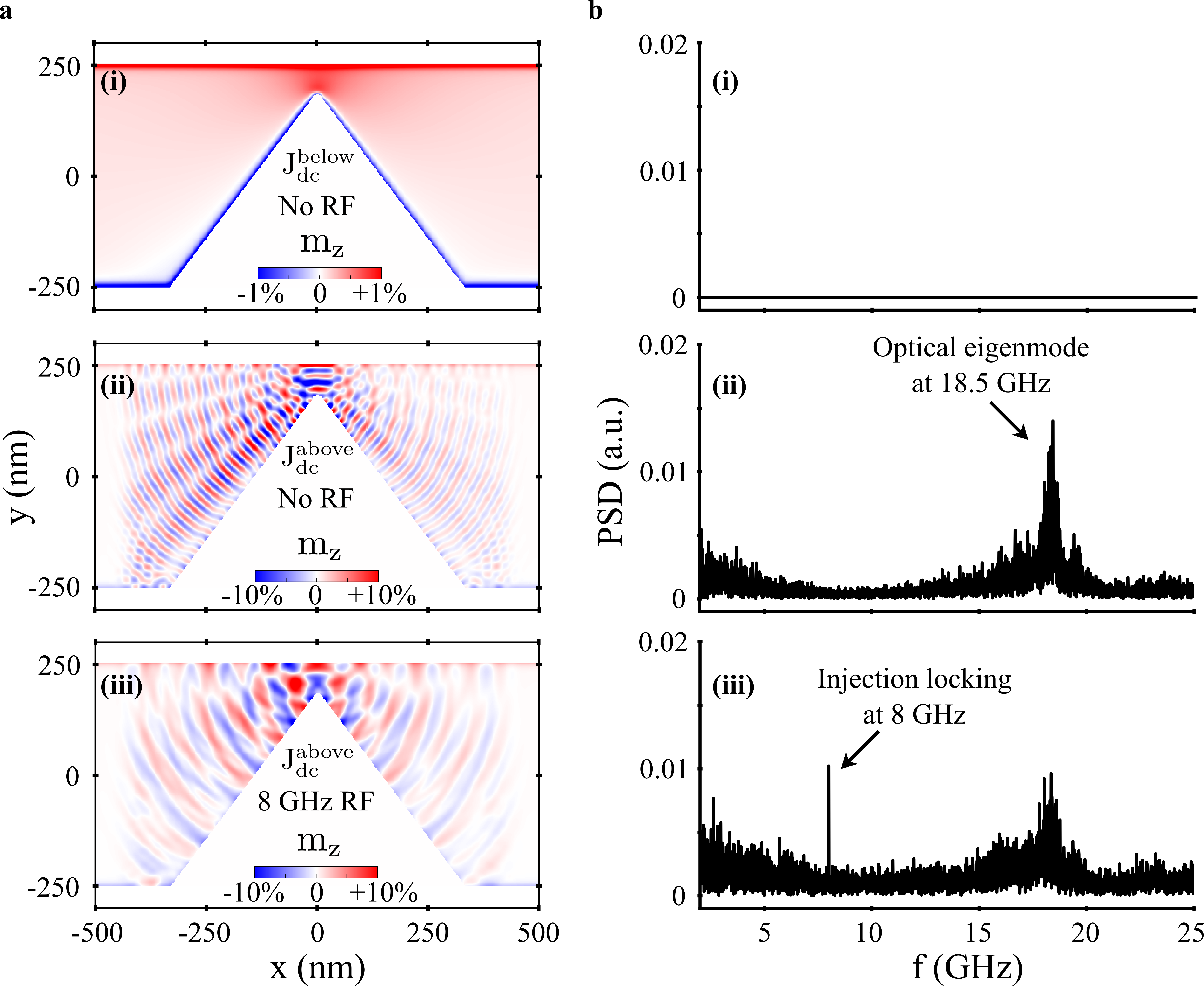}
\caption
{
}
\label{fig_2}
\end{figure}

\newpage
\begin{figure}[h!]
	\hspace*{-0cm}
	\includegraphics[scale = 0.95]{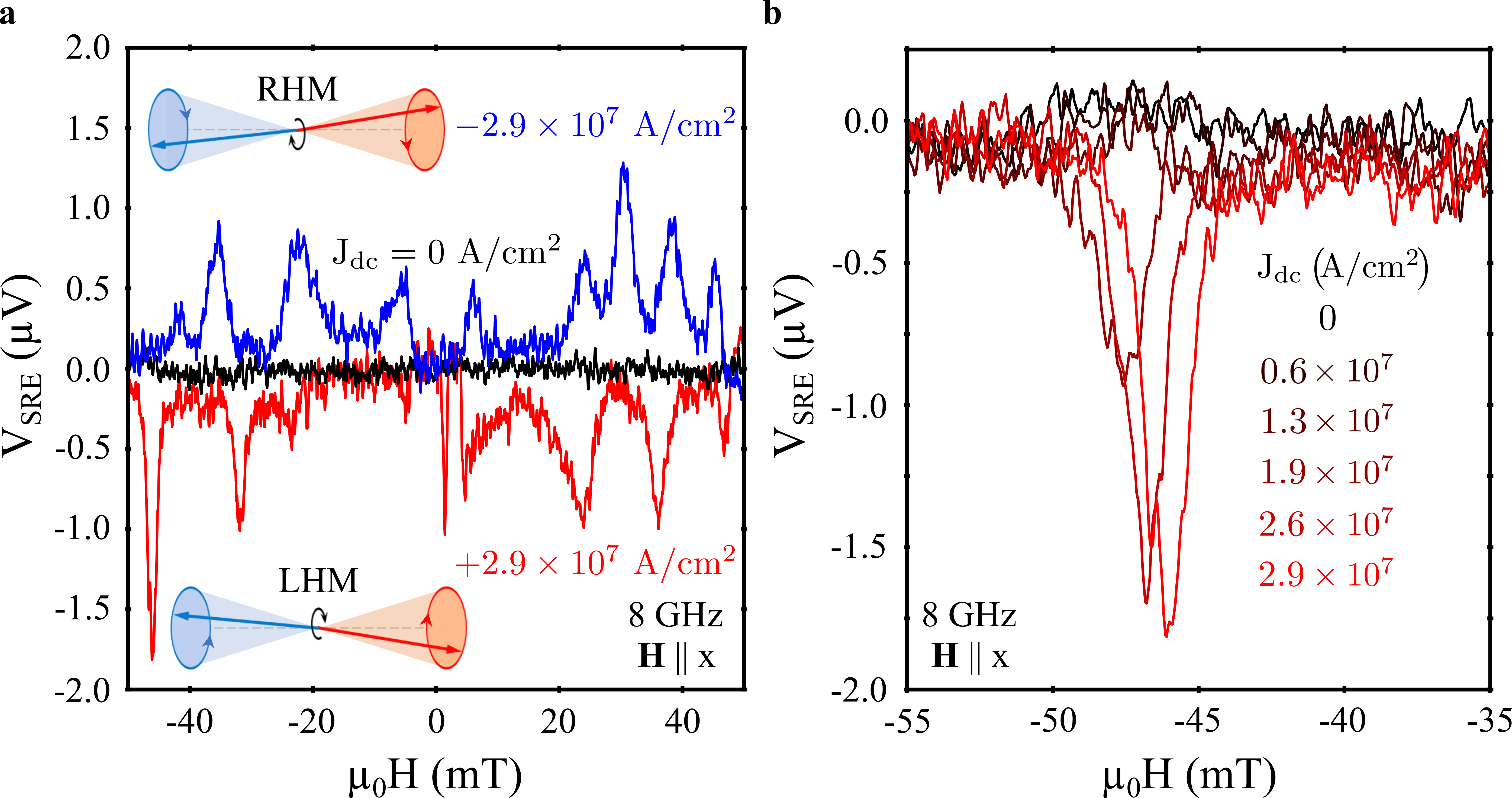}
	\caption
	{
	}
	\label{fig_3}
\end{figure}

\newpage
\begin{figure}[h!]
	\hspace*{-0cm}
	\includegraphics[scale = 0.95]{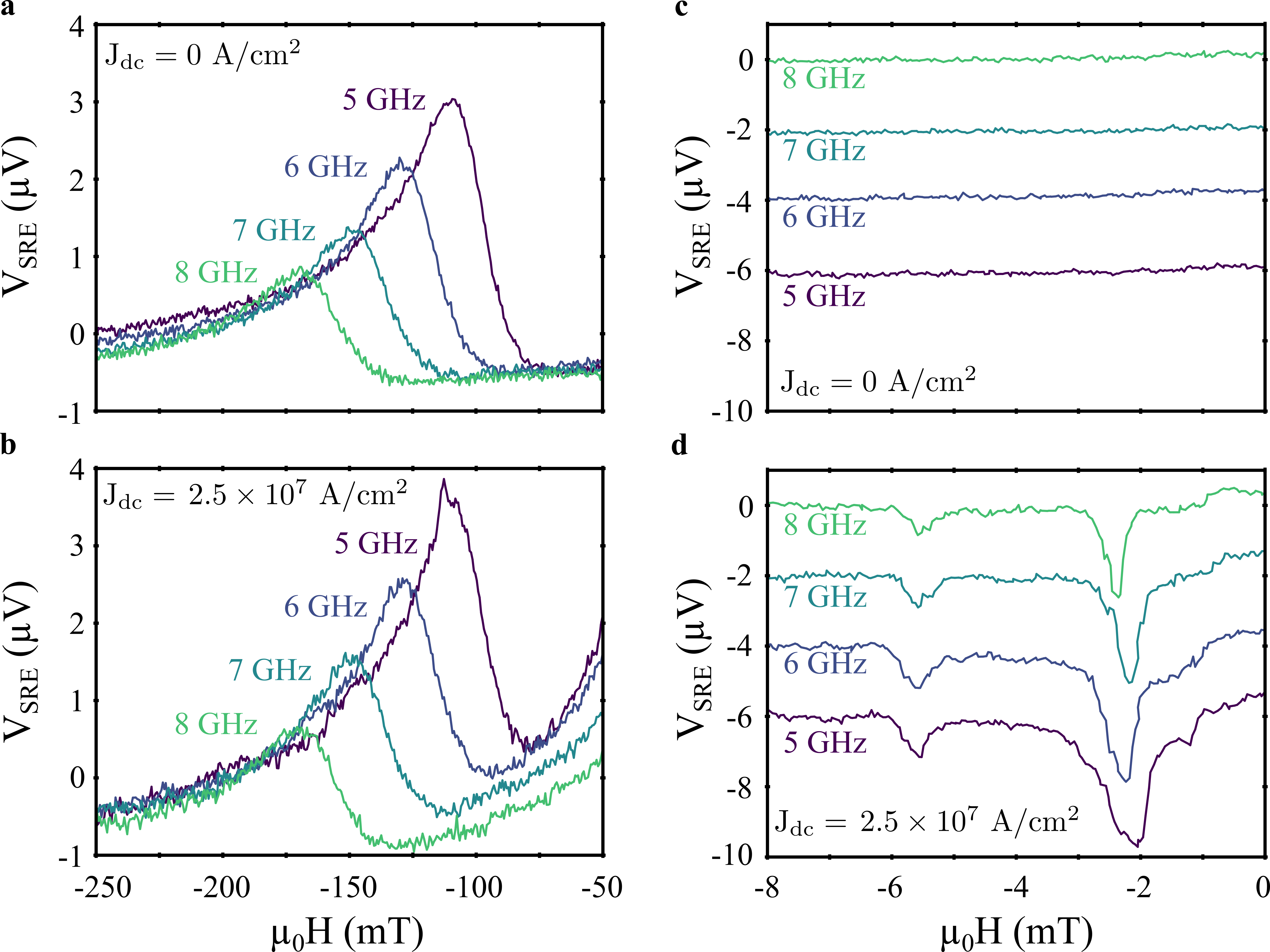}
	\caption
	{
	}
	\label{fig_4}
\end{figure}

\end{document}